\documentstyle[psfig]{l-aa}              
\def\hu{HU~Aqr}
\def\pe{$\phi_{\rm ecl}$}
\def\he2{He\,{\sc ii}\,$\lambda$4686}
\def\h1{He\,{\sc i}\,$\lambda$4471}
\def\kmps{km\,s$^{-1}$}

\begin{document}
\thesaurus {06(08.02.1, 08.02.2, 08.14.2, 08.13.1, 08.09.2 HU Aqr 
(=\,RX\,J2107.9--0518), 02.01.2)}
\title{
Phase-resolved high-resolution spectrophotometry
of the eclipsing polar HU Aquarii}
  \author {Axel D.~Schwope\inst{1,} 
              \thanks{Visiting astronomer,
              German-Spanish Astro\-no\-mi\-cal Center, Calar Alto, 
              operated by the Max-Planck-Institut f\"ur Astronomie, 
              Heidelberg, jointly with the Spanish National Commission 
              for Astronomy.}
	\and 
           Karl--Heinz Mantel\inst{2,*} 
	\and
	   Keith Horne\inst{3}
	}
   \offprints{Axel D.~Schwope}
                               
  \institute {
	Astrophysikalisches Institut Potsdam, An der Sternwarte 16,
	14482 Potsdam, FRG {\it (e-mail: ASchwope@aip.de)}
\and	Universit\"atssternwarte M\"unchen, Scheinerstrasse 1, 
	81679 M\"unchen, FRG
\and	Univ.~of St.~Andrews, School of Physics and Astronomy, 
	North Haugh, St.~Andrews, Fife KY16 9SS, Scotland, UK
}
\date{Received, accepted}
\maketitle
\markboth{Schwope, A.D. et al.: The eclipsing polar HU Aqr}{}
\begin{abstract}
We present phase-resolved spectroscopy of the bright, eclipsing polar \hu\
obtained with high time ($\sim$30\,sec) and spectral (1.6\AA ) resolution when 
the system was in a high accretion state. The trailed spectrograms reveal
clearly the presence of three different line components with different 
width and radial velocity variation. By means of Doppler tomography their
origin could be located unequivocally (a) on the secondary star, (b) the
ballistic part of the accretion stream (horizontal stream), and (c) the 
magnetically funnelled part of the stream (vertical stream). For the first
time we were able to derive a (near-)complete map of the stream in a polar.
We propose to use Doppler tomography of AM Herculis stars
as a new tool for the mass determination
of these binaries. This method, however, still needs to be calibrated by 
an independent method. The asymmetric light curve of the narrow 
emission line originating on the mass-donating companion star reveals 
evidence for significant shielding of 60\% 
of the leading hemisphere by the gas 
between the two stars.
\end{abstract}

\section{Introduction}
Polars or AM Herculis binaries are strongly magnetic cata\-clysmic 
binaries. The
white dwarf's magnetic field locks both stars in synchronous rotation and 
prevents the formation of an accretion disk. Instead mass transfer occurs
via an accretion stream which is thought to follow initially a ballistic 
trajectory and later governed by the magnetic field, which leads the matter 
along magnetic field lines to the polar regions of the white dwarf. Most of
the gravitational energy is released there, although some dissipative energy 
loss is expected to occur all along the stream, in particular in the coupling 
region where the stream is forced to leave the orbital plane. 

The stream manifests itself observationally in mainly three ways, (A) by 
so-called absorption dips observed in IR, optical or X-ray light curves
(provided the inclination is high enough and the observer and the accretion 
spot are (blue in color) 
located on the same side of the orbital plane), (B) by residual
emission during eclipses of the white dwarf, and (C) by 
strong, asymmetric and variable emission lines (H, He\,{\sc i}, He\,{\sc ii}
and other high-ionized species) at optical and UV wavelengths
(provided the system is in a high accretion state). The dips mentioned in (A)
occur, if the line-of-sight crosses the out-of-plane parts of the 
stream. They are caused by photoelectric (at X-ray wavelengths), grey or 
free-free (at optical/IR wavelengths) absorption and are richly structured,
indicating a highly fragmented stream. An overview of relevant data and
analysis was given recently by Watson (1995). 
Phenomenon (B) is less studied because only a handful of eclipsing polars 
are known and the numbers of photons 
collected at eclipse which certainly arise from the stream are rather few, 
hence, model\-ling of these events is widely missing in the literature, but
nice observational data meanwhile have been obtained e.g.~for UZ For (Schmidt 
et al.~1993), WW Hor (Beuermann et al.~1990) and HU Aqr (Schwope et al.~1995).
The emission lines (C) have attracted by far most attention because their 
occurence
is one of the `classical' identification criterion. A diverse 
appearance is reported in the literature, they may consist of broad, narrow,
high- and medium-velocity as well as (quasi-)stationary components, all 
with different radial velocity amplitudes and systemic velocities. These 
components may have a different appearance when a specific system is 
re-observed occasionally (see e.g.~the review of Mukai 1988). The mini\-mum 
distinction usually made discerns between a narrow line attributed to the 
secondary star and a broad underlying component whose origin is assumed 
to be somewhere in the accretion stream (see e.g.~Liebert \& Stockman 1985,
who have shown that for the systems known at that time the radial velocity 
amplitude of the narrow line lies between that expected for the $L_1$ and 
the center of mass of the companion). 
Later on, Rosen et al.~(1987)
have presented high-resolution spectra of V834\,Cen and distinguished 
4 subcomponents, which were thought to originate at different parts of the
accretion stream and the secondary star. Schwope (1991) and
Schwope et al.~(1991, 1993a) 
made use of repeated measurements of the narrow line in order to derive 
long-term ephemerides of the secondary star for
the non-eclipsing systems QQ\,Vul, MR\,Ser and V834\,Cen.
Recent attempts to locate the 
line emission in polars applied tomographic techniques to data obtained
on VV\,Pup and RX\,J0515.6+0105 (Diaz \& Steiner 1994, Shafter et al.~1995).

\hu\ is a system which has all of the observable features of the stream
mentioned, at least occasionally. It was discovered 
independently by British and German astronomers during their
optical identification programmes of ROSAT WFC and PSPC sources detected 
in the corresponding sky surveys (Schwope et al.~1993b, Hakala et al.~1993).
Since it was the brightest eclipsing polar with the most extended eclipse
at that time (meanwhile surpassed by RX\,J0515.6+0105) it attracted immediate
interest. We initiated an intensive optical follow-up study using
4 telescopes simultaneously and performed spectroscopy with high and low
spectral resolution, high-speed photometry simultaneously in UBVRI-colours
as well as standard one-channel photometry using a V-filter. All data
have good signal/noise-characteristics and provide a comprehensive 
database suitable to reach deeper insight into the kinematics and 
accretion processes of the whole class. Here we present as a first paper
in an upcoming row our results of high-resolution spectroscopy.

\section{Observations and reductions}
\hu\ was observed with the 3.5m-telescope and Cassegrain double-beam 
spectrograph 
(TWIN) at Calar Alto, Spain, on August 17/18, 1993. Both arms of this double 
beam spectrograph (beamsplitter at $\sim$5500\AA ) were equipped
with low-noise Tek-CCDs (1K, pixel size 24$\mu m$) as detectors. We used
a 1\farcs 8 wide slit thus matching the approximate size of the seeing disk.
The spectral coverage was 4180--5070\AA\ in the blue arm, 7450--8770\AA\ in 
the red arm. The red spectra were taken in order to search for the photospheric
Na-doublet from the secondary star. Since \hu\ was in a high accretion state
at that time this attempt failed. We, therefore, restrict our analysis to
the spectra taken in the blue arm. The grating used there with 1200 grooves per
mm yielded a FWHM spectral resolution of $\sim$1.6\AA\ in the center, and 
$\sim$3\AA\ at the edges of the spectra. 

\begin{figure}
\psfig{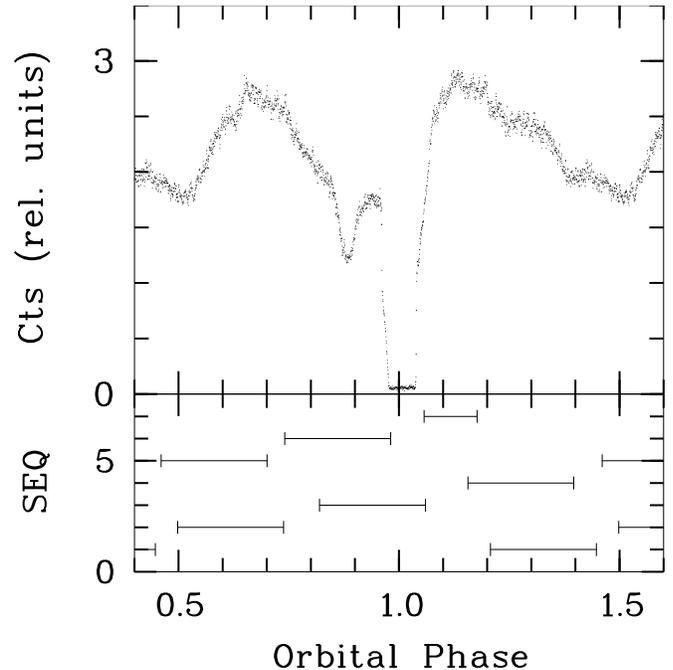}
\caption[meanlc]{\label{mean_lc}
{\it (upper panel)} Phase-averaged optical B-band lightcurve of HU Aqr 
for the period August 16-18, 1993. The counts of HU Aqr 
were reduced to that of 
a simultaneously observed nearby comparison star. Binsize of the lightcurve
is 3.75 sec (2000 phase bins).
{\it (lower panel)} Phase coverage of the individual trailed spectra of length
30 min (no. 1-6) and 12.5 min (no. 7).
}
\end{figure}

In order to avoid velocity smearing a radial velocity study of an AM~Herculis 
binary requires a phase
resolution of $\Delta \phi \simeq 0.05$ (Mukai 1988), hence $\sim$6$^m$
for a binary like \hu\ with period $P_{\rm orb} = 125^m$.
The time overhead between subsequent exposures at the TWIN, on the other hand,
is $\sim$5$^m$ due to CCD-readout, 
data transfer and storage of the large-format images from both 
arms of the spectrograph, which prevents a dense phase coverage. 
We resolved the time conflict by trailing the object along the slit 
while exposing. Individual exposure times were 30$^m$ with a
nominal trailing rate of 15 sec/pixel, which, taking into account the
size of the seeing disk of $\sim$2~pixels, finally yielded a time resolution
of $\sim$30 sec or 0.004 phase units. A total of 6 spectra with 1800\,s
and one with 750\,s integration time were obtained yielding full-phase
coverage.
Wavelength calibration of the flatfielded
and bias-subtracted two-dimensional images was performed
using Helium-Argon arc spectra taken every hour, a small shift of the arc line
pattern on the CCD due to flexure of the spectrograph was accounted for during 
reduction of the data. 
Cosmic-ray removal was achieved by median filtering small regions of size
$5\times 5$ pixels around suspected cosmics which were identified
by Laplace-filtering of the raw images.
Sky-subtraction was performed afterwards in the usual manner by fitting
polynomials to CCD-columns above and below the region exposed to the 
target star. 

\begin{figure*}[t]
\psfig{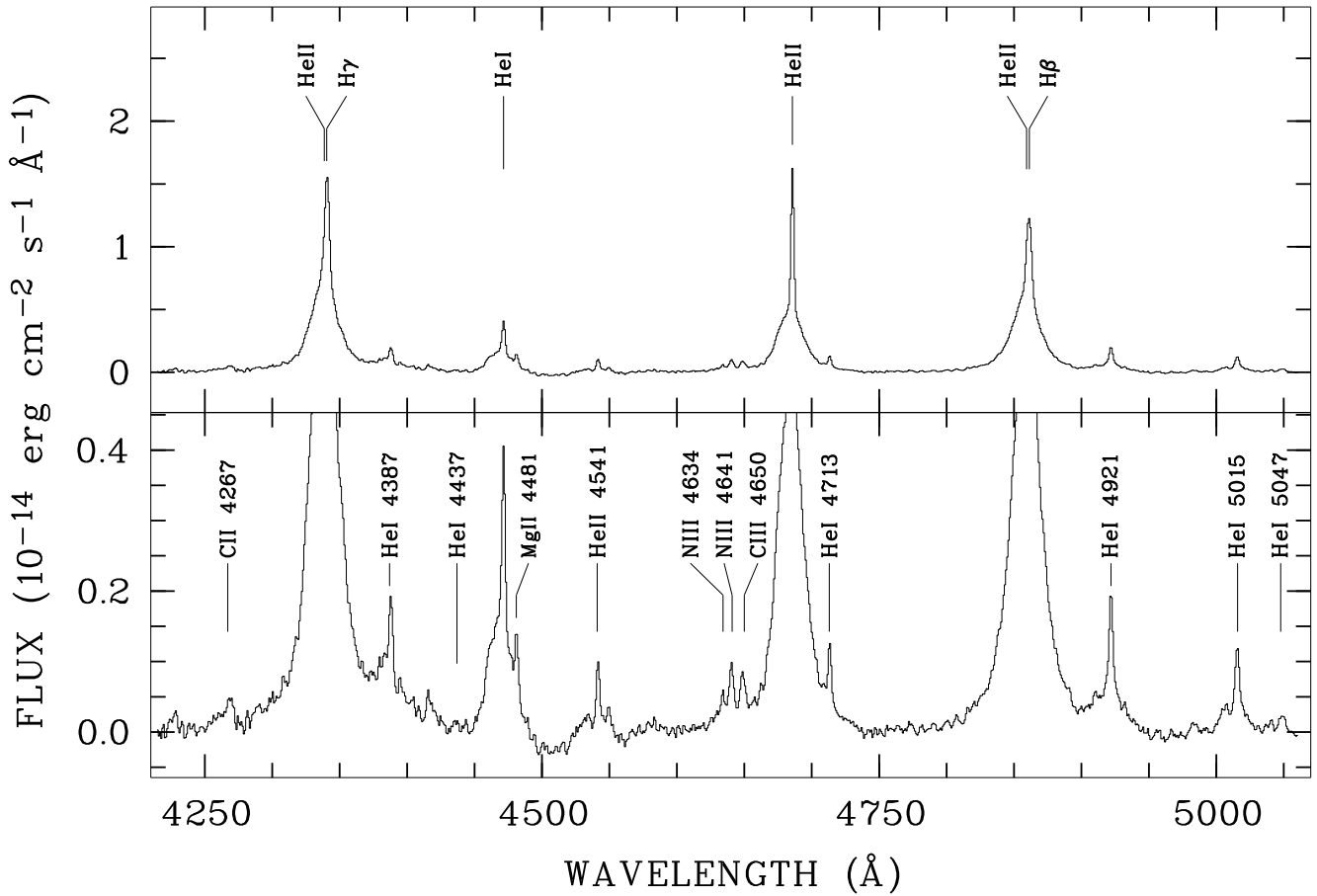}
\caption[meanspec]{\label{mean_spec}
Mean-orbital high-resolution spectrum of \hu\ on August 17, 1993, after 
subtraction of the continuum and after correction to zero velocity (narrow
emission  line). The same data are shown twice, in the bottom panel on 
an expanded intensity 
scale in order to facilitate identification of faint lines
}
\end{figure*}

Next, the spatial coordinate of individual pixels along the slit
had to be transformed to time or orbital phase. We defined start and end
of the exposure as coniciding with those pixels lying most nearly to half 
intensity between average sky- and (sky+object)-level in an adjacent
10 pixel surrounding. This resulted in a trailed spectrum with
116 or 117 pixels for an 1800 sec exposure, slightly different
from the 120 pixels expected from the nominal trailing rate of the telescope.
Since the lightcurves extracted from the trailed spectrograms
and the simultaneous photometry yielded no systematic offset, we assigned
the discrepancy to a slightly incorrect trailing rate of the telescope and 
estimate the timing accuracy of the spectrum in a single CCD-row to be better 
than 15 sec.

Simultaneously to our spectroscopy we obtained high-speed photometric data
(0.5 sec resolution) using the 2.2m-telescope also located at Calar Alto which 
was equipped with the MCCP (Barwig et al.~1987). The MCCP allows
monitoring of the target star, a compa\-rison star and the nearby sky 
simultaneously in UBVRI colours. In order to correct slit-losses 
and guiding errors of our trailed spectra, we first flux-calibrated them 
using observations of the white-dwarf standard star Wolf 1346, folded them 
through the response of the MCCP B-filter, binned the MCCP-data in 15 sec
time bins according to the timing of our trailed spectra and finally 
multiplied individual rows of the trailed spectra using the ratio of both light
curves (photometric, spectroscopic).

Final steps of data reduction included normalization of the spectra  
by subtracting low-order polynomial fits to the continuum and phase-folding
into 100 phase bins using our updated photometric ephemeris of eclipse center
\begin{equation}
T_{\rm ecl} = {\rm HJD} 244\,8896.543707(27) + E \times 0.086820446(9).
\end{equation}

\section{Results and analysis}
\subsection{The lightcurve}
In Fig.~\ref{mean_lc} (upper panel) we show the mean light curve in the B-band 
measured with the MCCP at the 2.2m-telescope. In the lower panel of the
diagram the phasing and phase coverage provided by our trailed spectra 
is displayed.
The most remarkable features of the light curve are: a) the deep eclipse 
of the white dwarf and the accretion spot by the late-type companion
star, b) a pronounced pre-eclipse dip centered on phase 
\pe = 0.88, c) a double-humped shape outside eclipse and dip, and d) 
strong QPO/flaring outside eclipse and dip. 
A detailed discussion of the light curves will be given elsewhere, in the 
context of our present analysis the interesting features are the eclipse 
and the dip. 
\begin{figure*}[th]
\psfig{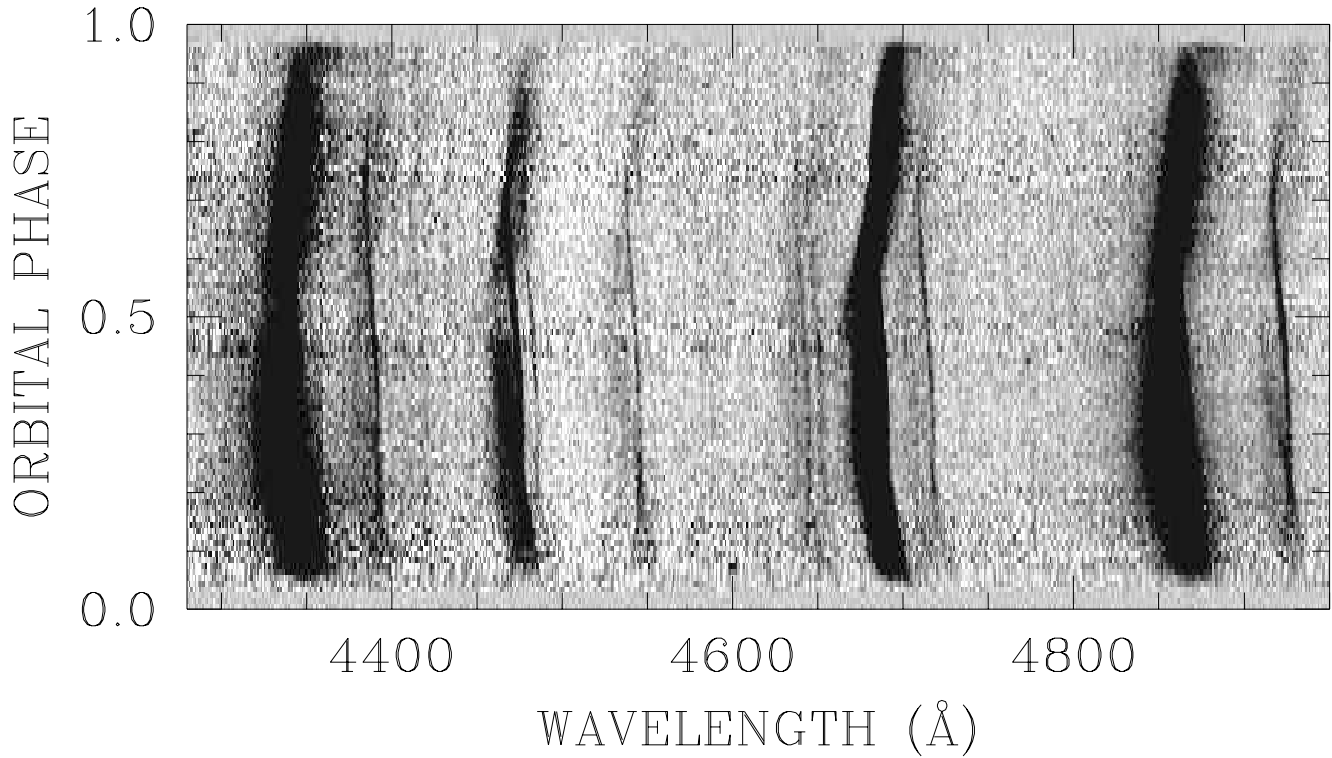}
\caption[trailed_spec]{\label{trailed_spec}
Trailed, continuum-subtracted, high-resolution spectrum of \hu\ on 
August 17, 1993, computed by phase-averaging the individual trailed spectra
using 100 phase bins. 
The dynamic range of this grey-scale image with wavelength running along
the abscissa and phase along the ordinate was set appropriately to 
emphasize weak lines.
The behaviour of the strong lines is better visible on the zoomed 
representations of Fig.~\ref{trailed_lines}
}
\end{figure*}

The eclipse is two-stepped with the first step being 
very steep ($\sim$3\,sec) followed by a more shallow second step with ingress
time of about 120\,sec. All four steps have different heights. We 
identify the obscured sources of light with the hot accretion plasma 
on the white dwarf emitting cyclotron radiation (steep ingress/egress) 
and the accretion stream (shallow ingress/egress). Both emitters radiate
anisotropically which gives the explanation for the different stepsizes 
at ingress and egress. The eclipse of the white dwarf itself is not 
clearly resolved in our data. 
We measured the length of the eclipse as the difference 
between the maxima in the first derivative 
of the lightcurve at ingress and egress
phase to be 584.6\,sec. This value represents the eclipse length of the 
accretion spot on the white dwarf surface. The spot is nearer to the secondary
star than the center of mass of the white dwarf CM$_{\rm wd}$. 
The eclipse of the CM$_{\rm wd}$ is shorter by a factor $\alpha_{\rm CM}/
\alpha_{\rm spot}$ with $\alpha$ being the opening angle of the secondary 
star as seen from the CM and the spot, respectively. The minimum value of 
this correction factor is 0.9792 ($\Delta t_{\rm ecl,CM} = 572.4$\,sec), 
calculated for the extreme (and therefore unlikely) assumptions of inclination 
$i = 90\degr$ and spot co-latitude $\beta = 0\degr$ (spot in the orbital 
plane). 

The Gaussian shaped dip is centered on \pe $= 0.882 \pm 0.003$ (deflection 
angle $d = 42.5\degr\pm 1.1\degr$ with respect to the line joining both stars)
and has a FWHM $=0.034\pm 0.001$ phase units. A $\pm 2\sigma$-range around
dip center (95\% of the absorbed flux) spans a range $d = 32\degr - 53\degr$.
The emission lines are not obscured at dip phase (see Figs.~\ref{trailed_spec}
and \ref{trailed_lines}), only the cyclotron continuum is absorbed. The dip
is therefore naturally explained as obscuration of the accretion spot 
by the transient accretion stream as in other polars (see Sect.~1). 
Since \hu\ is a system with a high inclination, dip center and width mark
exactly the location and extent of the coupling region in the orbital plane.

\subsection{The mean orbital spectrum}
In Fig.~\ref{mean_spec} we show the mean-orbital, continuum-subtracted,
radial-velocity-corrected spectrum of \hu. For radial-velocity correction
we used the $\sin$-fit to the narrow emission line of \he2 (see below). 
The profiles of the emission lines in Fig.~\ref{mean_spec} therefore show
a spiky narrow emission line atop broad line wings.
We can identify the emission lines normally found in AM Herculis binaries,
as e.g.~the Balmer lines H$\beta$ and H$\gamma$, the prominent lines of 
neutral (He\,{\sc i}\,$\lambda\lambda$4387, 4471, 4713, 4921, 5015) 
and ionized Helium (\he2), but find in addition several features 
not easily recognized in other systems. We note in particular the partially 
resolved C{\sc iii}/N{\sc iii}-structure at 4635-4650, 
C{\sc ii}\,$\lambda$4267, a second line of
ionized He (He{\sc ii}\,$\lambda$4541) with an additional unidentified 
fainter line in its red wing and Mg{\sc ii}\,$\lambda$4481 in the red wing
of He{\sc i}\,$\lambda$4471. 
\begin{figure*}[t]
\begin{center}
\begin{minipage}[]{85mm}
\psfig{figure=fig4_hgam_ps,width=85mm,bbllx=35mm,bblly=83mm,bburx=181mm,bbury=236mm,clip=}
\end{minipage}
\begin{minipage}[b]{9mm}
\end{minipage}
\begin{minipage}[]{85mm}
\psfig{figure=fig4_he1_ps,width=85mm,bbllx=35mm,bblly=83mm,bburx=181mm,bbury=236mm,clip=}
\end{minipage}

\vspace{0.5cm}
\begin{minipage}[]{85mm}
\psfig{figure=fig4_he2_ps,width=85mm,bbllx=35mm,bblly=83mm,bburx=181mm,bbury=236mm,clip=}
\end{minipage}
\begin{minipage}[b]{9mm}
\end{minipage}
\begin{minipage}[]{85mm}
\psfig{figure=fig4_hbet_ps,width=85mm,bbllx=35mm,bblly=83mm,bburx=181mm,bbury=236mm,clip=}
\end{minipage}
\end{center}
\caption[trailed_lines]{\label{trailed_lines}
Grey-scale representation of trailed, continuum-subtracted, high-resolution 
spectra of the strong emission lines of \hu\ on 
August 17, 1993 (zoomed representation of Fig.~\ref{trailed_spec}). 
Phase runs along the ordinate from bottom to top, wavelength has been 
transformed to velocity using the rest wavelengths of the specified lines.
The dynamic range for all representations is for all lines approximately
between 0 and 80\% of maximum intensity.}
\end{figure*}

The C{\sc iii}/N{\sc iii}-complex is resolved into three lines, a single line
of N{\sc iii} ($\lambda 4634.2$) and two blends of N{\sc iii} 
($\lambda\lambda 4640.6, 4641.9$) 
and C{\sc iii} ($\lambda\lambda 4647.4, 4650.3, 4651.5$). The lines of C and N
are of comparable strength. According to Williams \& Ferguson (1983) the 
N{\sc iii}-lines are probably excited by continuum fluorescense. This 
does not apply to C{\sc iii}\,$\lambda$4650 or to 
C{\sc ii}\,$\lambda$4267 which do not couple directly to the ground states
via electric dipole resonance lines. Whereas C{\sc ii}\,$\lambda$4267 may 
be excited by dielectronic recombination no final conclusion was 
reached by Williams \& Ferguson 
about the probable excitation mechanism of the C{\sc iii}-lines, apart
from the general statement, that `some selective excitation mechanism is 
operative for this line'. We would like to note simply here that 
this special excitation 
mechanism does not require other conditions regarding
density, temperature and external radiation field than are required for 
excitation of 
the other transitions of the high-ionization species 
found in the spectrum of \hu, since the kinematics and structure of
all high-ionization lines seem to be the same.

Three lines of the He{\sc ii}-Pickering series ($n=4$) 
lie in our spectral window at
$\lambda\lambda$4859.32 ($m=8$), 4541.59 (9), and 4338.67 (10). The line
at 4541\AA\ is clearly resolved whereas the other two blend with 
H$\beta$ and H$\gamma$. The wavelength-difference between the H-Balmer 
and the He{\sc ii}-lines is 2.01\AA\ and 
1.80\AA , respectively, which is at the limit or below our spectral 
resolution at the specified wavelengths. From our low-resolution spectra
we estimate the flux ratio between the non- or weakly-blended
lines at 4541.59\AA\ ($m=9$) and 5411.52\AA\ ($m=7$), $F_{5411}/F_{4541} \simeq
1.9$, and interpolate to estimate the contribution of He{\sc ii}\,$\lambda$4859
to H$\beta$, which is of the order of 10\%, but variable as a function 
of phase.  H$\gamma$ itself is stronger than H$\beta$ and the blending 
He{\sc ii}-line ($m=10)$ is weaker than that at H$\beta$, hence we expect 
a much weaker contamination of H$\gamma$.


The Balmer lines show an inverted Balmer decrement with 
$I(H\beta )/I(H\gamma ) = 0.86$. The value for optically thin `case B'
recombination is $\sim$2.12 (Pottasch 1984), 
hence the Balmer lines are significantly optically
thick. Although also He{\sc ii}\,$\lambda$4686 is not optically thin (see 
below), the fact that this line is only weakly
blended (by He{\sc i}\,$\lambda$4713 and C{\sc iii}\,$\lambda$4650) and that
our spectral resolution is highest just at this line makes it the most 
suitable one in our spectra for detailed investigations, but before doing so we
take a general look at the trailed spectrograms. 

\subsection{Trailed spectra}
In Figs.~\ref{trailed_spec} and \ref{trailed_lines} we show the 
phased-averaged trailed spectra in overview (weak lines emphasized) and for
the strong lines H$\beta$, H$\gamma$, \h1, \he2\
in detail with a phase resolution of 100 phase bins.
In Fig.~\ref{trailed_spec} it can be seen that the 
phase intervals centered on \pe = 0.11, 0.46, and 0.78 
which were covered only once during our observations
have higher noise in the continuum, the effect on the S/N in the
strong lines is nevertheless negligible.
All emission lines have the same structure and can be subdivided
into three emission line components (best visible in \he2). Most striking is
a narrow emission line (henceforth referred to as NEL), which is most 
prominent around \pe$ = 0.5$ and which has zero radial velocity at conjunction 
of the secondary star. This component is naturally explained as being 
of reprocessed origin from the illuminated hemisphere of the secondary star,
seen here in unprecedented clearness. The second component has a more transient
nature, it is a relatively narrow line, moving in a zig-zag manner from 
maximum recessional velocity ($>1000$\,\kmps ) around eclipse phase to large
negative velocities ($<-1500$\,\kmps ) around \pe = 0.45. This component 
is brightest in the phase intervals \pe$ = 0.20-0.25$ and $0.70-0.75$, around 
those phases where it crosses the NEL. We refer to this component as 
high-velocity component HVC. The third component is a broad underlying 
one visible throughout the whole orbital cycle (outside eclipse), with
maximum blueshift around \pe$ \simeq 0.40$ which we refer to as broad base
component BBC. The HVC and the BBC must originate in the stream.
\begin{figure}[th]
\psfig{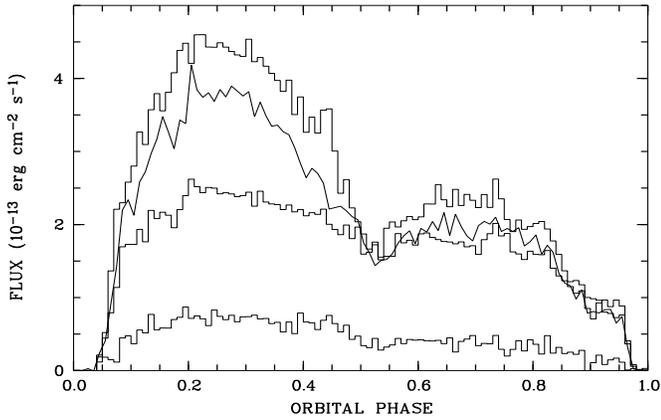}
\caption[line_lcs]{\label{line_lcs}
Variation of the integrated line flux of the four main emission lines
as a function of phase. Shown are from top to bottom the light curves for 
H$\gamma$ (histogram), H$\beta$ (polygon), \he2 , and \h1. 
}
\end{figure}

Although the four main emission lines have the same general appearance
differences exist in details. This concerns the width of the NEL, 
which is broader in the H-Balmer lines than in the He-lines, and 
the lightcurves of the lines (brightness variation as a function of phase).
The width of the NEL of the different lines is addressed in the 
next section (3.4); the lightcurves are shown in Fig.~\ref{line_lcs} 
(integrated flux over the wavelength interval used for display
in Fig.~\ref{trailed_lines}). The intensity of the stream components
dips to a minimum at phase \pe $\simeq 0.53$, and is generally stronger 
before than after this dip. This asymmetry is generally stronger for 
the low-ionization lines H$\gamma$, H$\beta$, \h1 with flux ratios
$F(\phi < 0.53)/F(\phi > 0.53) = 2.09, 2.14, 2.02$, respectively, 
than for the high-ionization
\he2-line with the corresponding flux ratio of 1.54.

A likely explanation for the brightness variation of the HVC, which is 
pronounced only at certain phases, 
lies in the illumination geometry and significant optical thickness of the 
stream in the low-ionization lines. The Balmer lines and \h1
are bright when the stream is seen from its concave, that is its 
illuminated side
(see Fig.~\ref{stream_geo} for a sketch of the geometry). The convex,
non-illuminated side of the stream visible after \pe$ = 0.53$
lies in its own shadow and is therefore much fainter than in the less
optically thick \he2-line.

\subsection{The width of the NEL in He- and H-lines}
The trailed spectrograms of the Balmer lines (in particular their NELs)
look somewhat blurred or diffuse when compared with the He-lines. 
Here we investigate, if this is caused by the weak blending lines of
the He{\sc ii}-Pickering series mentioned above. 
The mean width (full width at half maximum FWHM) of the four strongest 
emission lines (H$\gamma$, \h1, \he2, H$\beta$) in the phase interval
$=0.50-0.56$ (NELs intensive, stream components relatively faint) is  
$(4.19\pm0.18)$\AA, $(2.74\pm0.15)$\AA, $(2.47\pm0.07)$\AA, and 
$(4.34\pm0.12)$\AA\, respectively. 
The spectral resolution of arc lines at the wavelengths of our lines 
is 2.33\AA, 1.92\AA, 1.67\AA, and 1.76\AA, which yields intrinsic widths 
of the NEL
of 3.48\AA, 1.95\AA, 1.82\AA, and 3.97\AA. The He-lines have the same 
widths within the errors (1.95 vs.~1.82\AA ), whereas the Balmer lines
are significantly broader. 

We have synthesized blended line profiles of H$\beta$ plus 
He{\sc ii}$\lambda 4859$ assuming the same intrinsic line width for both 
lines of 1.82\AA \ (that of the \he2 -line) and a maximum relative intensity 
of 20\% of the He-line. After folding with the instrumental resolution at 
H$\beta$, the resulting unresolved line had a FWHM of 3.8\AA \
(observed 4.34\AA ). We conclude that the NELs of the Balmer lines are
intrinsically broader than those of the He-lines. Using the same scheme
of synthesizing line profiles, we could estimate the width of the H$\beta$-NEL
to be $\ga 3.5$\AA . Below we show that the width of the He{\sc ii}-line
is explained by velocity broadening introduced by the range
 of orbital velocities over the illuminated hemisphere of the
secondary star. Hence, each line which originates in that region should have
at least this width. If we assume that the observed excess width of H$\beta$
is caused by thermal Doppler broadening, we would need 
temperatures in excess of 
$2\times 10^6$\,K in order to reflect the observations. This temperature,
corresponding to coronal temperatures, is unbelievably high for a line 
originating in a (quasi-)chromosphere. The canonical temperature for the
Balmer-line emission region in active M-stars is $10^4$\,K and the observed
line widths at H$\alpha$ are about $\la 2$\,\AA \ (Giampapa et al.~1982, 
Worden et al.~1981). This is also true for irradiated atmospheres
of dMe stars although our knowledge of this very complex process is very 
poor (see e.g.~Cram 1982).
A more likely explanation seems to be that the observed
line width reflects orbital motion of a more extended Balmer line region,
perhaps some kind of wind/mass outflow component that co-rotates with the 
secondary star. An even simpler possibility is `saturation broadening' due 
to large optical depths in the Balmer lines.

\subsection{Gaussian deconvolution of \he2 }
The conventional approach to investigate the emission line structure
in polars is fitting of Gaussians. Due to the lack 
of spectral and temporal resolution often a distinction between 
a broad and a narrow component is made and only two Gaussians are used 
but sometimes up to four (Rosen et al.~1987). 
We followed that approach and performed a triple-gaussian fit for the NEL,
the HVC and the BBC in an iterative fashion. 

\begin{figure}[th]
\psfig{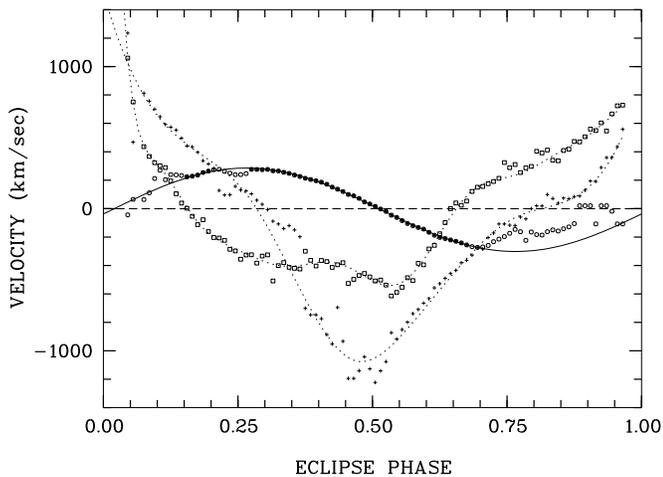}
\caption[vrad]{\label{vrad}
Radial velocities of the three subcomponents of \he2, determined
by fitting Gaussians with fixed widths to the phase-binned spectra 
shown in Fig.~\ref{trailed_lines} (NEL: open and filled circles, HVC:
crosses, BBC: squares). The solid line is a $\sin$-fit to the NEL-data 
shown with filled symbols, the dashed lines are smooth interpolations
for the HVC and BBC used as guess-values for subsequent iterations
}
\end{figure}

In a first step we fitted three Gaussians to each of the phase-binned 
spectra shown in Fig.~\ref{trailed_lines} while fixing the widths (FWHM) 
of the NEL and the HVC at 2.5\AA and 4.5\AA, respectively. This gave
very good estimates for the radial velocities, and good estimates for the 
line intensities and the FWHM of the BBC. The latter displayed no distinct 
orbital variation, instead it showed an erratic scatter of 1.5-2\,\AA \ around
$11.5$\,\AA . Results for the radial velocities thus obtained are shown 
in Fig.~\ref{vrad}. The fits gave unreliable results for the line intensities 
and for the wavelengths at those phases where two components are crossing 
each other. This happens at phases \pe$\sim$0.23 and $\sim$0.70 
between NEL and HVC and at \pe$\sim$0.35 between HVC and BBC. In order to 
provide input guess values for a second iteration,
the phase-dependent radial
velocities were approximated by smooth curves, a sinusoid for the NEL, 
spline-curves for the HVC and BBC. 
The sinusoid and the splines are shown as solid and dashed curves in 
Fig.~\ref{vrad}, the sine curve was calculated using only the data points
drawn with filled symbols. 
In a second step we fitted all individual 15 sec spectra in the trailed 
spectrograms, where these fits were restricted to a small
range around the interpolated guess values and 
the NEL was fixed on the input guess values.
Similarly, the phase-binned trailed spectrogram was fitted 
a second time using these smoothed guess values from the first step but
allowing the FWHM to vary freely within certain ranges. 

\begin{figure}[th]
\psfig{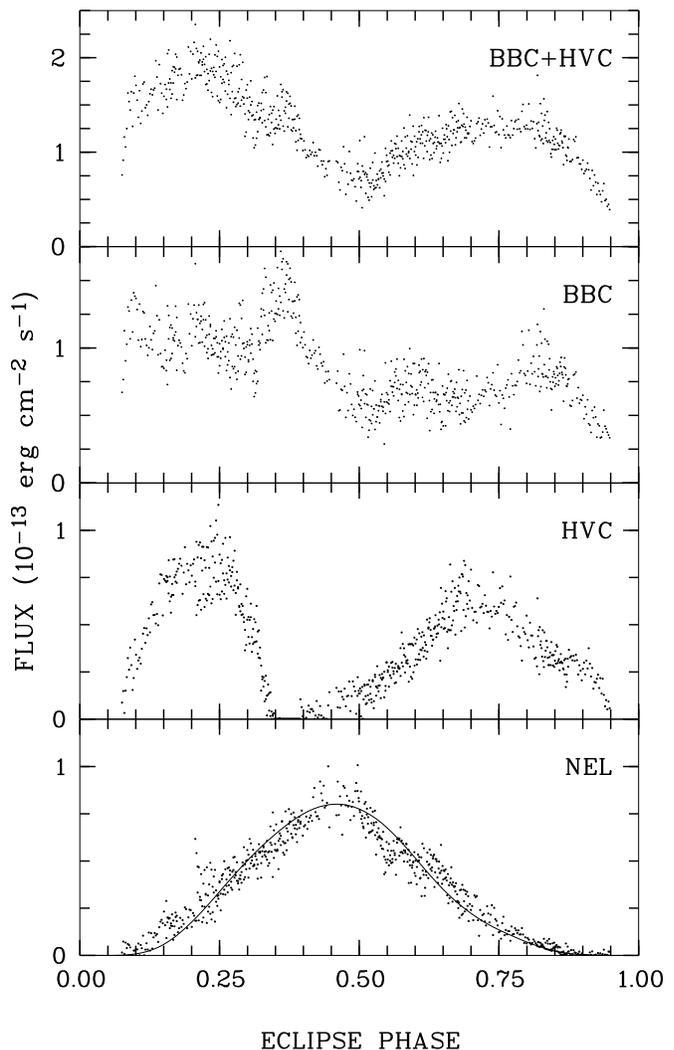}
\caption[lc_3comp]{\label{lc_3comp}
Light curves of the three subcomponents of \he2, determined
by Gaussian fits to individual trailed spectra as described in the 
text. The uppermost panel represents the summed light from the accretion 
stream, the lowest that from the secondary star. The solid line in the 
lowest panel is the result of a model calculation, described in Sect.~3.6
}
\end{figure}

The results of these subsequent steps of the fitting procedure,
the intensity variations of the three emission line components and the
variation of the FWHM of the NEL, are shown in Figs.~\ref{lc_3comp} and
\ref{fwhm_nel}. In Fig.~\ref{fwhm_nel} we show only those data, for which 
the NEL
could be clearly separated from the HVC. In Fig.~\ref{lc_3comp} we also show 
the summed lightcurve of HVC and BBC representing the light from the 
stream. 
The lightcurve of the BBC can be roughly described as a two-stepped function,
with constant brightness before and after superior conjunction
of the secondary. It is brighter by $\sim$50\% during first half of the 
orbital cycle than during the second half. The hump centered on \pe$\simeq 
0.35$ is an artefact of the fitting procedure, the excess light there is
missing in the HVC-lightcurve, the same may apply to the less pronounced
hump at phase 0.82. The lightcurve of the HVC has two pronounced humps
separated by $\sim$0.5 in phase with the first hump centered on 
\pe$\simeq$0.22. The sum of both, the HVC and the BBC, shows two smooth humps.

The lightcurve of the NEL finally is single-humped. This 
component is brightest around superior conjunction of the secondary, but 
has a remarkable asymmetry with respect to phase 0.5. It is brighter
during first part of the orbital cycle, with maximum brightness  
at \pe = 0.46. In some of our fitting experiments, depending on chosen 
input values and caging of parameters, the brightness of the NEL
became even compatible with zero at phases later than 0.75 but it has 
$\sim$20\% of the peak flux at the symmetric phase \pe$ = 0.25$.

The width (FWHM) of the NEL is largest at and slightly after superior 
conjunction (\pe$ = 0.5$), reaching
a maximum value of about 2.45\AA . We were prevented from really measuring
the width for phases later than $\sim$0.63 due to the decreasing brightness 
of the NEL and interference with the other line components. 

The radial velocity variation  of the NEL is in excellent agreement with a 
$\sin$ curve. The fit assuming a circular orbit for the data points shown
with filled symbols in Fig.~\ref{vrad}, $v(\phi) = v_{\rm sys} + K^\prime_2
\sin (2\pi (\phi - \phi_0))$, yields $v_{\rm sys} = -7.8\pm 1.4$\,\kmps ,
$K^\prime_2 = 293.5 \pm 1.3$\,\kmps , and $\phi_0 = 0.016 \pm 0.001$.
There is a small additional systematic uncertainty not larger than 
$\sim$3\,\kmps for $K^\prime_2$ dependent on the data points included in the 
fit. If we correct the systemic velocity to the solar system barycenter,
$\Delta v = -3.5$\,\kmps, it becomes $v_{\rm sys} = -11.3 \pm 1.4$\,\kmps .
We refer to the radial velocity amplitude as $K^\prime_2$ (instead
of $K_2$) because it represents the velocity of the center of light 
of the illuminated hemisphere instead of the center of mass. 

Glenn et 
al.~(1994) have presented some 10 high-resolution spectra of H$\alpha$
obtained when the system probably was in a somewhat different state
of activity. Probably due to the lack of phase-resolution they discern 
between a broad and a narrow component only and derive $v_{\rm sys} = 83\pm 
44$\,\kmps, $K = 244 \pm 44$\,\kmps, and $\phi_0 = 0.04 \pm 0.04$, 
respectively, for their NEL. 
The results of both studies are compatible with each other 
only at the 2$\sigma$-level, hence we cannot completely rule out real
changes of the emission line structure. 
Glenn et al.~have described the broad component
of H$\alpha$ also with a sinusoid. Given our higher-quality data we cannot
use such a simple approach. 

Both of our stream components show maximum 
recessional velocity around eclipse, the full amplitude of the HVC is 
$\sim$2000\,\kmps, that of the BBC $\sim$1400\,\kmps. 
Neither the HVC nor the BBC display a radial
velocity curve which can be described by a simple mathematical function.
An  analysis of these components
requires a different approach, which is presented below as tomogram analysis.
If one, nevertheless, e.g.~for comparison with the Glenn et al.~analysis,
fits a sinusoid to the radial velocity data of the BBC, $v(\phi ) = \gamma
+ K \sin (2\pi (\phi - \phi_0 ))$, one gets $\gamma = 3 \pm 10$\,\kmps,
$K = 534 \pm 15$\,\kmps, and $\phi_0 = -0.32 \pm 0.01$. Again, these values
are compatible with the Glenn et al.~analysis at the 1--2$\sigma$ level.
One question which is left open here in particular is, 
whether the two components HVC and BBC originating in the stream
really reflect emission from two separate regions, as it appears, 
or whether instead the emission all along the stream naturally produces 
skewed profiles 
simply due to optical depth, projection and curvature of the stream.
The conventional approach of deconvolution using Gaussian fits, however,
reveals very important information for the interpretation of the tomograms
as far as e.g.~intensity variation is concerned. 
 
The asymmetrical shape of the lightcurve of the NEL, the non-zero
velocity at eclipse and the non-vanishing phase offset $\phi_0$ are 
pronounced and require an interpretation. It appears likely that all
three features are related to the same origin, an asymmetric illumination 
of the secondary star.

\subsection{A reprocessing model for the NEL}

The skewness of the lightcurve of the NEL is significant. The sole explanation
for such an asymmetry is to assume an asymmetric illumination of the secondary
star. Similar asymmetries, although mirrored with respect to the axis of 
symmetry, have been observed by Beuermann \& Thomas (1990) in the 
nova-like UX UMa-system IX Vel and by Hessman (1989) in the dwarf nova IP Peg. 
The likely mechanism for the asymmetry there
is significant irradiation from the hot spot where the accretion stream
impacts on the disk. Since this lies on the leading side of the secondary
the trailing side is less irradiated and consequently fainter. Such an
explanation cannot work in an AM Her binary, since the source of irradiation 
is bound to the surface of the white dwarf. 

A related type of asymmetry has been discovered in NaI absorption lines
of AM Her itself by Southwell et al.~(1995) and Davey \& Smith (1996).
These authors detect flux deficits on the trailing side of the secondary 
in the NaI absorption lines caused by stronger heating of the trailing 
hemisphere, and explain the necessary shielding of the leading side
by absorption in the accretion column near the white dwarf, which points
to the leading side in AM Her. 

We have synthesized line profiles originating from the illuminated 
hemisphere of the secondary star using the method as outlined in 
Horne \& Schneider (1989) and Beuermann \& Thomas (1990). In this model 
the Roche lobe surface is covered by a set of discrete surface elements. The 
flux incident on each surface element is calculated assuming a point source 
located at the position of the white dwarf. Lightcurves and line profiles
are synthesized by summing contributions from all surface elements visible
at a given phase. At this stage we allowed for optically thin emission, 
being proportional to the incident flux, as well as for optically thick
emission, where this flux was weighted with the projected surface of the 
element in concern. No further line broadening mechanism than velocity 
broadening over the illuminated Roche surface and the instrumental broadening 
were taken into account. 
We calculated a series of these profiles as a function 
of phase and shielding. We took the simple view of describing shielding
by a weighting factor for each surface element of the leading side 
which was constant for the whole hemisphere. We performed such calculations for
three values of the mass ratio $Q = M_{\rm wd}/M_2 = 3, 4, 5$ assuming that
the secondary follows the ZAMS mass-radius relationship given by 
Caillault \& Patterson (1990) (in that case the mass of the secondary is 
$M_2 = 0.146$\,M$_\odot$ and only very weakly dependent on mass ratio)
and for $Q=2.5, M_2 = 0.35$\,M$_\odot$, as motivated by our tomogram 
analysis. The orbital inclination $i$ was chosen according to the 
$q-i$-relation for the observed length of the eclipse (see Fig.~\ref{qi_rel}) 
as $i = 80\degr, 82\degr, 85\degr, 90\degr$, respectively, for $Q = 2.5, 3, 
4,$ and 5.

\begin{figure}[th]
\psfig{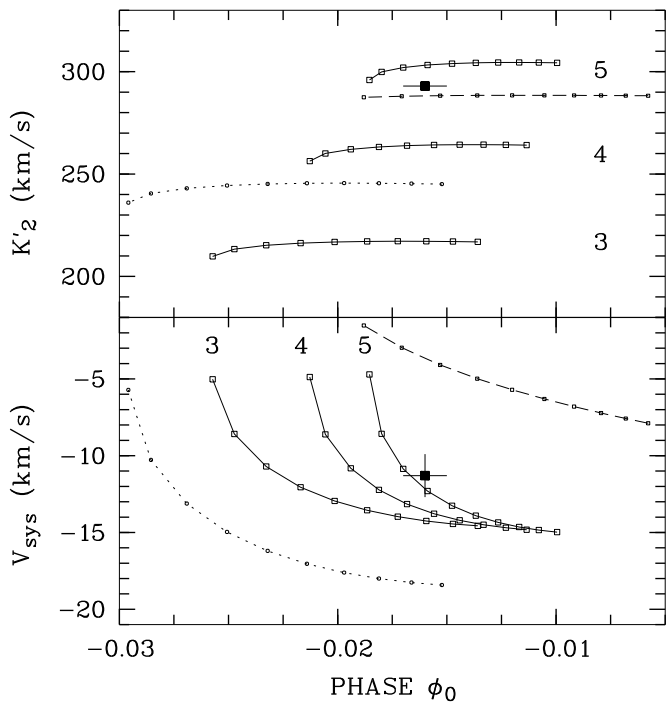}
\caption[nel_k2]{\label{nel_k2}
Results of model calculations for the NEL: Shown are the relation between
radial velocity amplitude, systemic velocity and phase shift obtained from
sine fits to synthetic NEL-profiles as a function of mass ratio and 
shielding of the leading hemisphere of the secondary star. The numbers
designate the mass ratio $Q = M_{\rm wd}/M_2$.
The large open
squares connected with solid lines are data computed in the optically thick 
mode, the small open squares connected with dashed lines are data computed 
in the optically thin mode (shown only for $Q = 5$) and the small circles
are optically thin data for $Q = 2.5$ with a massive secondary of $M_2 =
0.35$\,M$_\odot$. Shielding increases by 10\% between the data points starting 
with no shielding at right. The large filled squares are observed data.
}
\end{figure}

\begin{figure}[th]
\psfig{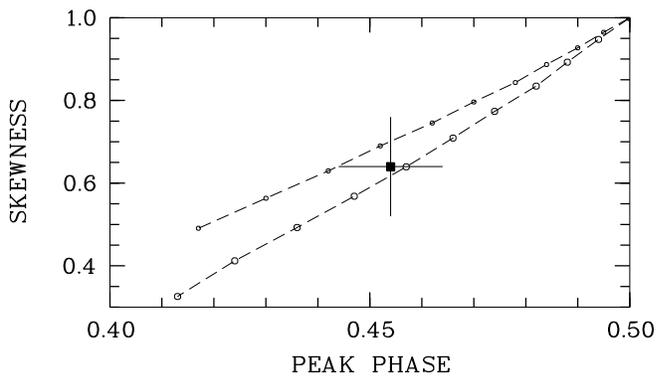}
\caption[nel_skew]{\label{nel_skew}
Results of model calculations for the NEL: Shown are the relation between
skewness ($I(\phi > 0.5) / I(\phi < 0.5)$) and phase of peak intensity
of the NEL-lightcurve for a mass ratio $Q=5$ and varaible shielding of the
secondary star. As in Fig.~\ref{nel_k2}, shielding grows from right to left
in 10\%-steps and the larger symbols represent optically thick models. The 
filled square is the observed value with estimated uncertainties.
}
\end{figure}

\begin{figure}[th]
\psfig{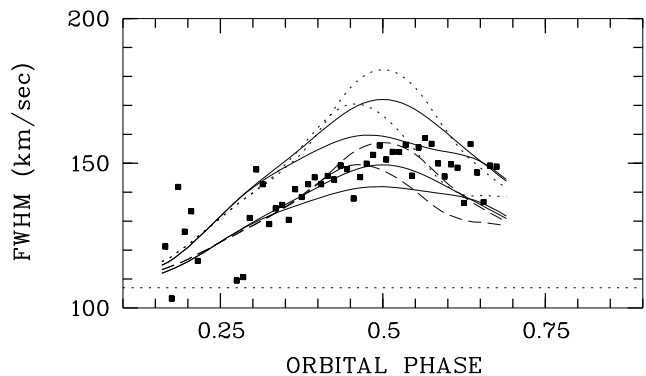}
\caption[fwhm_nel]{\label{fwhm_nel}
Full width at half maximum FWHM of the NEL 
transformed to velocity at the \he2 -line as a function of phase.
The filled squares are observed data, the lines are model curves calculated
assuming different mass ratios, mass of secondary star, optical thickness 
and shielding (see text for details). The dotted line at bottom indicates
the instrumental resolution.
}
\end{figure}

The synthesized profiles were subsequently analyzed in the same manner 
as our observed spectra but first their spectral resolution was degraded 
to a FWHM = 107\,\kmps as observed.
Then single Gaussians were fitted to determine the central wavelength. 
Afterwards sine curves were fitted to the modelled radial velocity data, 
taking into account only data from those phase intervals which were used
also to determine the observed radial velocity curve.  The results
of this procedure are shown in Figs.~\ref{nel_k2} (parameters of radial 
velocity curves) and \ref{fwhm_nel} (FWHM vs.~phase)
together with the observed data.
The modelled radial velocity curves are not exact sinusoids, but the deviation
is negligible for the phase intervals in concern.
Two families of model curves are shown in Fig.~\ref{fwhm_nel}. The upper
four curves are calculated for $Q=2.5, M_2=0.35$\,M$_\odot$, the lower four
for  $Q=5, M_2=0.146$\,M$_\odot$ (the other models with $M_2= 0.146$\,M$_\odot$
but smaller $Q$ are undistinguishable from that with $Q=5$). 
The solid lines are optically thick models,
the dashed and dotted lines optically thin. The upper curve of each pair
is for no shielding, the lower curve for 60\% shielding.

In addition to the line profiles lightcurves for the reprocessed lines 
as a function of phase were generated. They show a distinct asymmetry 
in the required manner, which was parametrized using (a) 
a skewness defined as ratio of fluxes $F(\phi > 0.5) / F(\phi 
< 0.5)$ and (b) the phase of maximum intensity. Since these parameters
are negligible functions of the mass ratio (they depend only on the adopted 
shielding), we show in Fig.~\ref{nel_skew} only one set of model data for 
$Q = 5$ and optically thin and thick modes, respectively, together with 
the parameters derived from the observed lightcurve of the NEL. 

The models and data shown in Figs.~\ref{nel_k2} and \ref{nel_skew} 
suggest consistently a mass ratio $Q \simeq 5$, optically thick emission 
and an effective shielding of the leading side of the secondary star by 
60\%. Lightcurves for the optically thick and thin models were shown also
by Horne \& Schneider (1989), where it can be seen that the thin models
predict a plateau around \pe = 0.5. This happens when always the same numbers
of surface elements are visible for a distinct phase interval. Such a plateau
is not observed in HU Aqr and we conclude from all the detailed comparisons
between our models and the observed data, that the NEL is non-negligibly
optically thick. 

The fit is not completely satisfactory for the variation 
of the FWHM. While the low-$Q$, high-$M_2$ model predicts too high
values for the FWHM for most orbital phases (upper four curves) the model
with $Q=5$ predicts somewhat too small values, at least if one takes shielding
into account and uses the optically thick models. In addition, shielding
moves the phase of maximum FWHM from \pe = 0.5 towards smaller values
whereas the observed FWHM peaks only at \pe$ \simeq 0.5-0.55$. 
We note, however,
that the FWHM is the least well determined quantity among all 
those used here for comparison between observation and models and we 
do not regard the deviation between observation and simulation 
as a severe problem.

\subsection{Tomogram analysis of the four main emission lines}

In this section we investigate the trailed spectrograms shown in 
Fig.~\ref{trailed_lines} using a 
tomographic inversion technique known as filtered backprojection and
described e.g.~by Robinson et al.~(1993). Making use of such an inversion 
technique implies that any emission not observed at rest wavelength of a
line is regarded as shifted due to the Doppler effect and the backprojection
from the observers $(v,\phi )$-space to the original $(v_x, v_y)$-plane 
thus makes use of the whole line profile (instead making use of 
only e.g.~width, intensity
or center). The interpretation of such maps is mostly done by comparison 
with computer-generated schematic drawings of the possible locations
of line emission in the binary using as input parameters only the velocites 
of the two stars of the binary.

The computation of a predicative Doppler-tomogram (or Doppler-map or shortly 
map)
requires a large number of 
projections (high phase resolution) and a high spectral resolution. 
Our dataset, in particular the trailed spectrogram of \he2, 
is unique in this respect. The tomograms shown as grey-scale images
in Fig.~\ref{tom_grey} and as contourplots in Fig.~\ref{tom_cont} have a
velocity sampling of 5\,\kmps \ for \he2 \ and 20\,\kmps \ for the other lines.
Two features in the maps are striking, a bright spot at velocities
$v_x \simeq 0-30$\,\kmps, $v_y \simeq 290$\,\kmps \ and
 a cometary tail linked to it and extending 
to $v_x \la -1000$\,\kmps\ at $v_y \simeq 0$\,\kmps. A third feature appears
as faint smeared blotch in the lower left quadrant of the 
maps. Since this blotch has a low contrast in Fig.~\ref{tom_grey} only we show
in Fig.~\ref{tom_stream} the Doppler map of those emission components of
\he2 \
attributed to the accretion stream. It was computed by subtracting the 
Doppler image of the NEL from that of the complete profile and shows
two main features, the cometary tail and the blotch below it.

\begin{figure*}[t]
\begin{center}
\begin{minipage}[]{85mm}
\psfig{figure=fin_hg_q4_ps,width=85mm,bbllx=34mm,bblly=94mm,bburx=194mm,bbury=249mm,clip=}
\end{minipage}
\begin{minipage}[b]{9mm}
\end{minipage}
\begin{minipage}[]{85mm}
\psfig{figure=fin_he1_q4_ps,width=85mm,bbllx=34mm,bblly=94mm,bburx=194mm,bbury=249mm,clip=}
\end{minipage}

\vspace{0.5cm}
\begin{minipage}[]{85mm}
\psfig{figure=fin_he2_q4_ps_new,width=85mm,bbllx=34mm,bblly=94mm,bburx=194mm,bbury=249mm,clip=}
\end{minipage}
\begin{minipage}[b]{9mm}
\end{minipage}
\begin{minipage}[]{85mm}
\psfig{figure=fin_hb_q4_ps,width=85mm,bbllx=34mm,bblly=94mm,bburx=194mm,bbury=249mm,clip=}
\end{minipage}
\end{center}
\caption[tom_grey]{\label{tom_grey}
Doppler maps of the four main emission lines of \hu\ computed by 
filtered backprojection of the trailed spectra shown in 
Fig.~\ref{trailed_lines}. The velocity resolution for the H$\gamma$-, 
He{\sc i}-,
He{\sc ii}-, and H$\beta$-lines is 161, 129, 107, and 109\,\kmps, respectively,
at the origin and 173, 143, 124, 126\,\kmps \ at $v = 1000$\,\kmps.
The overlay shows the shapes of the secondary 
star and the accretion stream in Doppler coordinates for an assumed 
mass ratio of $Q = M_{\rm wd}/M_2 = 4$. 
}
\end{figure*}

Backprojection transforms sine curves to points, the bright spot
thus corresponds to the NEL and represents the irradiated secondary star. 
For the \he2-line we compare the position of the spot in the Doppler map
with the results of our Gaussian fits. The spot appears at $v_x = 30$\,\kmps
and $v_y = 283$\,\kmps .
The 
offset of the spot with respect to the symmetry axis $v_x = 0.0$ reflects the
obscuration of the leading (left) side. The ratio of the velocity components
corresponds to the measured phase offset $\phi_{\rm 0,sin}$ 
of the radial velocity curve $\phi_{\rm 0,dop} = (2\pi)^{-1} 
\arctan (v_x/v_y) = 0.017$, 
in good agreement with $\phi_{\rm 0,sin} = -0.016$. The velocity amplitude 
$K^\prime_{\rm 2,dop} = \sqrt{v_x^2 + v_y^2} = 285$, 
however, is lower than found from the sine fit to the 
radial velocity curve $K^\prime_{\rm 2,sin} = 293.5$\,\kmps. This might be 
caused by deviations of the true line profiles from Gaussian curves. We see
such deviations pronounced in our modelled line profiles in the optically thin 
approximation and they are present too but less pronounced in the optically 
thick models (for an illustration see Horne \& Schneider, 1989, their 
Fig.~12). We have checked the hypothesis 
by comparing radial velocities measured (a) by sine fitting  
radial velocity curves and (b) by 2D-Gaussian fitting in tomograms of our 
modelled profiles. 
We indeed find differences of the required type 
of the order 3--8\,\kmps depending on mass ratio 
and optically thin/thick approximation. An additional small bias towards
lower velocities of the observed peak in the tomogram is likely due to the 
underlying cometary tail. So we regard both methods to be in 
agreement. The maps calculated for our synthesized NEL-profiles show
that even with our low spectral resolution (with respect to the width of the 
NEL which is not well sampled) we should have seen a triangle-like structure
in the Doppler map sketching the heated `nose' of the secondary star if
the emission happens in the optically thin  approximation. In the optically 
thick approximation the modelled tomogram yields only a slightly elongated 
structure, very similar to what is observed. 

\begin{figure}[th]
\psfig{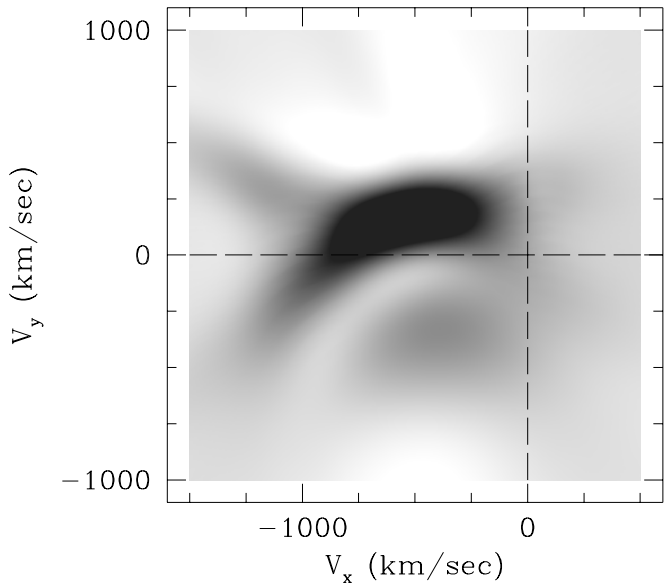}
\caption[tom_stream]{\label{tom_stream}
Doppler map of the stream components of \he2 computed by sutracting the
Doppler image of the NEL from the Doppler image of the whole line 
profile. In order to emphasize the  blotch at $(v_x,v_y) \simeq 
(-400,-300)$\,\kmps, we adopted a linear intensity
scale between 5\% and 60\% of the peak intensity.
}
\end{figure}

The Roche lobes shown as overlays in Fig.~\ref{tom_grey} have all the same 
size thus facilitating a comparison of the four emission lines. The 
Roche lobes were calculated for a mass ratio of $Q = M_{\rm wd}/M_2 = 4$.
According to the larger widths of the NELs of the Balmer lines discussed 
in Sect.~3.4 the corresponding spots are more smeared than those of the 
Helium lines. In addition, emission is more concentrated towards the 
equator of the Roche lobe in the Balmer lines. All spots except that in the 
map of H$\beta$ are bright on the trailing side. The H$\beta$-spot has 
maximum intensity at $v_x \simeq 0.0$\,\kmps \ and a longish extension 
inclined by 135\degr\ towards negative $(v_x,v_y)$-velocities. The radial 
velocity amplitudes of the NELs of \h1, H$\gamma$ and H$\beta$ as measured
in the tomograms are $v = 284, 313, 312$\,\kmps, respectively. The value for 
\h1 \ agrees with that of \he2, both values for the Balmer lines agree with 
each other but are significantly higher than those for the Helium lines.

The cometary tail connected to the secondary star in the tomogram is the 
horizontal accretion stream. It is the Doppler image of the HVC 
as can be seen directly by mapping only this component singled out using 
the Gaussian fits. Again this feature appears most brilliant in the 
map of \he2. Its location and curvature
in the map are sensitive to the velocities of the stars in the binary and, 
hence, to the masses 
and below we will give mass estimates relying on the 
assumption that the ridge of light exactly reflects the location of the 
trajectory of the stream (approximated in the usual manner as 
single particle trajectory, Lubow \& Shu 1975, freely falling in the 
gravitional potential).

\begin{figure*}[ht]
\begin{center}
\begin{minipage}{87mm}
\psfig{figure=he2_cont_q25_ps,width=80mm,bbllx=40mm,bburx=198mm,bblly=97mm,bbury=251mm,clip=}
\end{minipage}
\begin{minipage}{87mm}
\psfig{figure=hg_cont_q5_ps,width=80mm,bbllx=40mm,bburx=198mm,bblly=97mm,bbury=251mm,clip=}
\end{minipage}
\end{center}
\caption[tom_cont]{\label{tom_cont}
Contour plots of the Doppler maps for \he2\ (left) and H$\gamma$ (right)
shown in Fig.~\ref{tom_grey} with 11 equidistant contours between 0\% and
100\% of the peak intensity.
The rhombus at $(v_x, v_y) \sim (-400,-300)$\,\kmps\ indicates 
the center of BBC emission, the small circle and 
the cross give the loci of the centers of mass of the white dwarf and the 
secondary star for the assumed mass ratios of $Q = 2.5$ (left diagram) and 5 
(right diagram). Shown are the
shape of the secondary star, the ballistic trajectory (horizontal stream,
upper left quadrant) and the part of the stream which has left the orbital 
plane (vertical stream, lower left quadrant).
}
\end{figure*}

The third component in the Doppler-map, the diffuse blotch in the lower left 
quadrant, can be identified with the BBC in the trailed spectrogram.
This again became evident by mapping the BBC alone which was isolated 
using the Gaussian fits (the corresponding center of emission of the BBC
is marked by a rhombus in Figs.~\ref{tom_grey} and \ref{tom_cont}). 
The blotch represents emission from the vertical stream. 

\begin{figure}[t]
\psfig{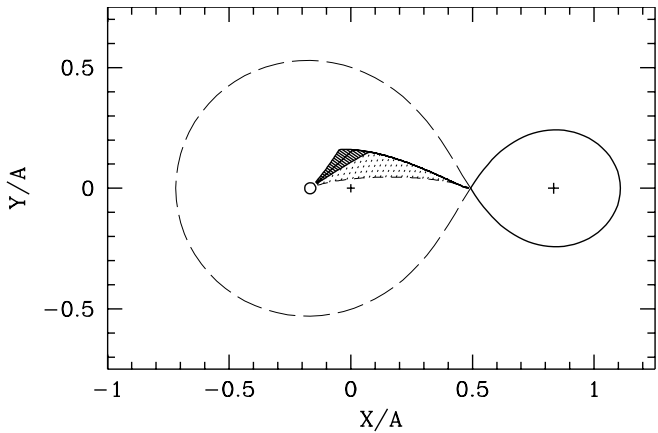}
\caption[stream_geo]{\label{stream_geo}
Scheme of the accretion geometry in \hu \ projected onto the orbital plane for
an assumed mass ratio $Q = 5$. Shown are the loci of the center of mass 
(cross), the white dwarf (circle) and the secondary star (cross at right)
as well as the Roche-lobes of both stars. Matter leaves the secondary star
at the $L_1-$point and follows a ballistic trajectory. Three field lines 
connecting the stagnation region and the white dwarf are shown, the region
in between is densily dotted. One additional field line (dashed line) connects
the inner Lagrangian point with the white dwarf. The region between this line 
and the dense vertical stream, where the accretion curtain is located,
is dotted lightly.
}
\end{figure}

\begin{figure}[th]
\psfig{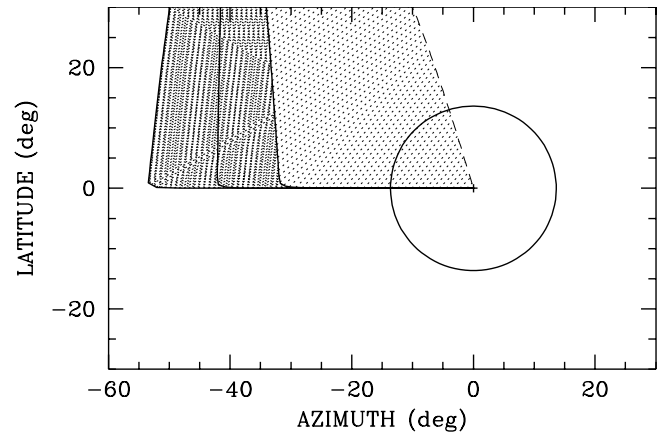}
\caption[view_fwd]{\label{view_fwd}
Scheme of the accretion geometry in \hu \ as seen by a hypothetical observer 
located at the center of the white dwarf. 
Matter leaves the secondary star
at the $L_1-$point and follows a ballistic trajectory, hence it is deflected 
under the influence of Coriolis forces to the left. The stream leaves the 
orbital plane and couples onto field lines at an azimuth of 43\degr, where the
absorption dip occurs. Three field lines 
connecting the stagnation region and the white dwarf are shown indicating 
the width of the absorption dip (dark dotted region).
One additional trajectory (dashed line) is shown leaving the plane at the 
$L_1-$point. 
The region between this line 
and the dense vertical stream is less dense dotted (accretion curtain).
}
\end{figure}

We have modelled the location of all components in the tomogram in a  
manner described e.g.~by Marsh \& Horne (1988) but included in addition a 
magnetic drag influencing the trajectory of the stream. The results are
shown for two extreme cases of the mass ratio in Fig.~\ref{tom_cont} ($Q=2.5$
and 5, respectively) and for an intermediate value of $Q=4$ in 
Fig.~\ref{tom_grey}. The
corresponding picture in positional coordinates is shown in 
Fig.~\ref{stream_geo}. In Fig.~\ref{view_fwd} we show the geometry of the
stream as seen by a hypothetic observer at the white 
dwarf. 

The shape of the secondary star projected onto the orbital plane is the 
same in position and Doppler coordinates, though rotated by 
90\degr\ in the Doppler coordinates because the orbital velocity 
vectors are perpendicular to the position vector. 
Our stream starts at the inner Lagrangian point $L_1$ with no initial
width. 
The stream then follows a single particle trajectory (Lubow \& Shu 1975) 
in the 
Roche-potential down to the interaction region between magnetosphere and 
accretion stream. The location and size of this stagnation region 
is determined from the phase and the phase width 
of the absorption dip (Fig.~\ref{mean_lc}). 
In Figs.~\ref{tom_cont} to \ref{view_fwd}
we show three trajectories with turn-off points adjusted to reflect the
center and width of the stream (distance to center is $\pm 2\sigma$ of 
Gaussian fitted dip). There the stream leaves the orbital plane 
as it couples onto field lines. We assumed no dissipation to occur in the 
interaction region. The abrupt re-orientation of the stream 
happens in a small spatial region, but produces a large jump in velocity from 
$\sim$(-1100,0) to $\sim$(-400,-300) in $(v_x,v_y)$-space. The stream then 
follows the field lines of an assumed dipolar field. 
The field axis was fixed at 
colatitude $\beta  = 25\degr$ measured with respect to the rotation axis
and azimuth $\Psi = 40\degr$ with respect to the line joining both stars. 
The former value derives from the observed motion of cyclotron lines in 
low-resolution spectra (Schwope et al.~1996, in preparation), the latter one
from the phase of the optical/X-ray bright phase center. 

We have chosen a model with $Q = 4$ ($K_1 = 87$\,\kmps, $K_2 = 350$\,\kmps) 
for the overlay in Fig.~\ref{tom_grey} which gives a fairly good 
representation of the four lines and their subcomponents regarded there. 
We note, however, that the HVC-emission of the \he2-, \h1-, and H$\beta$-lines
is concentrated below the line defined by the model, 
whereas that of the H$\gamma$-line
is concentrated above the model line. We find excellent agreement between 
the HVC-ridge of \he2 and the trajectory for a mass ratio of 
$Q = 2.5$ ($K_1 = 147$\,\kmps, $K_2 = 367$\,\kmps) 
and good agreement for H$\gamma$ using a mass ratio of $Q = 5$ 
($K_1 = 77$\,\kmps, $K_2 = 384$\,\kmps) (Fig.~\ref{tom_cont}).

We see in the tomogram that 
significant broadening takes place all along the horizontal stream, which is 
not reflected by our model. This is due to the simplicity of the model which 
does not take into account any potential broadening mechanism as e.g.~the
finite width of the stream and its optical thickness.

It is
somewhat surprising that we cannot detect emission from the stream at really 
high velocities ($> 700$\,\kmps) close to the white dwarf ($v_{\rm ff} \simeq 
4000$\,\kmps). This may reflect the effects of a high degree of ionization 
and small radiating surface of the stream at high velocities. 
Possible effects which in addition may contribute to the
low velocities in the BBC (and its Doppler image, the blotch) 
are (a) a slightly incorrect 
continuum subtraction thus removing faint emission at high velocities, 
and (b) emission not only from the accretion stream
but also from the accretion curtain,
hence from matter leaving the plane all along the horizontal stream. This 
matter is
thought to be responsible for shadowing of the leading side of the secondary
star and its re-emission has not been regarded so far. The turn-off points
of such trajectories lie closer to the $L_1$ and the jumps from the HVC- to
the BBC-region in the map are smaller thus closing the expected gap.
One sees also that the observed large width of the BBC is introduced naturally
by the models due to the simple fact that coupling happens along the 
trajectory over a significant distance and the jumps from the HVC- to the 
BBC-region adjust correspondingly. In principle, so-called `gamma-smearing'
may contribute to the observed width of the BBC in the Doppler map (Diaz \&
Steiner 1994). Gas with velocity perpendicular to the orbital plane $v_z$
produces a shift $\gamma = v_z \cos i$. The effect of this is to blur the
corresponding feature in the Doppler map over a ring of radius $\gamma$
centered at its usual position. If we assume $v_z = 1000$\,\kmps and 
$i$ in the range 81\degr--85\degr, we get a smearing amplitude of $156 
- 87$\,\kmps, which is of the order of or below the resolution of our
map. Hence, our map of \hu \ is essentially blind to the vertical motions.

In Figs.~\ref{stream_geo} and \ref{view_fwd} the distribution of matter 
in the system is shown. The dark shaded region in both figures are the 
horizontal and vertical streams, while the dotted region gives a schematic 
of the accretion curtain necessary for shielding the leading hemisphere.
This curtain is constrained by a trajectory which couples to the magnetic field
at the $L_1$-point. This gives not the required 60\% derived above
obscuration indicating 
that the view taken here is still too simplistic. Additional obscuration 
will be provided by nature by the finite width of the stream at the $L_1$
which is of the order of 10\% of the secondary star's radius.

We can use our stream models to obtain estimates for physical 
parameters of the stagnation region. The extent of this region in the 
orbital plane (length along the trajectory) is $l \simeq 6 \times 10^9$\,cm and
along which the undisturbed dipole field strength varies between 3 and 9 kG. 
The polar field strength
was set to $B = 38$\,MG which yields a field strength in the accretion spot
of about 36-37\,MG as required by the identification of cyclotron harmonics
(Schwope et al.~1993, Glenn et al.~1994). Our code assumes a constant cross
section along the trajectory, $\rho v = \dot{m} = \rm{constant}$. Coupling at 
different loci along the trajectory is possible by a variation of the 
density $\rho$, thus varying the ram pressure $p_{\rm ram} \propto \rho v^2$
of the stream. Under these assumptions we found that the 
ratio of the ram and magnetic pressure varies by a factor of 6--7 along the
trajectory in the stagnation region. 
The size of the stagnation region 
is transferred down to the white dwarf surface via dipole field lines where
it occupies a region with full opening angle of about 5\degr --6\degr 
corresponding to a linear dimension of $l \simeq 7 \times 10^7$\,cm
for a white dwarf with radius $R_{\rm wd} = 7 \times 10^8$\,cm.

The tomogram together with the schemes of the accretion geometry 
also yields hints for an understanding of the lightcurves of the 
HVC and the BBC (Fig.~\ref{lc_3comp}). Both components are brighter during
first half of the orbital cycle, when we view the magnetic curtain's 
concave illuminated side. 
The HVC is bright and easily to recognize in the trailed spectrogram 
around quadrature phases. The observed spectra at these phases correspond
to projections of the map parallel to the $v_x-$axis, hence along the 
cometary tail, producing a sharp feature in a spectrogram.
The observed spectra at conjunction phases, on the other hand, correspond 
to projections along the $v_y-$axis. At these phases emission from the 
horizontal stream is smeared over an interval of $\sim$1000\,\kmps \
which produces only very broad spectral features. Hence, at theses phases 
the HVC apparently vanishes and we obtain the minima in the lightcurve
of the HVC.

\section{Discussion and conclusions}

\begin{figure}[t]
\psfig{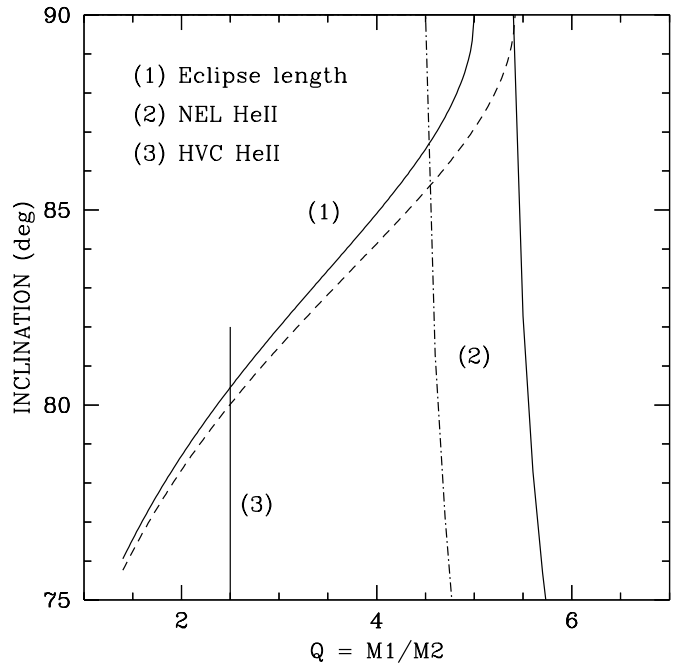}
\caption[qi_rel]{\label{qi_rel}
Mass determination of \hu \ using the observed length of the eclipse (1), the 
radial velocity amplitude of the illuminated hemisphere of the secondary 
star with different mass-radius relations (2), and the curvature of the 
ballistic stream in the Doppler map (3).
}
\end{figure}

We have found two distinct features in the trailed spectrograms of the four
main emission lines in the blue spectral regime of \hu, in particular the 
line of ionized Helium \he2, which are potentially useful 
for a determination of the stellar masses. These are the amplitude of the
radial velocity curve of the narrow emission line NEL and the location and 
curvature of the
horizontal stream in the tomogram, the Doppler image of the HVC. 
An additional constraint on the 
mass ratio is set by the observed eclipse length. All these constraints are
shown in a $(Q, i)-$plane in Fig.~\ref{qi_rel} with $Q$ being the mass ratio
and $i$ the orbital inclination. The eclipse length is dependent
on the size of the secondary star's Roche lobe and the inclination, hence
relation (1) in the diagram is purely geometrical (it is of course implicitly
assumed that the secondary fills exactly its Roche lobe) (Chanan et 
al.~1976). The solid line is valid for the measured eclipse length,
the nearby dashed line is valid for the maximum correction
which accounts for the offset of the accretion spot from the white dwarfs 
center, see above. Relation (2) 
in the diagram for the NEL makes use of a factor 
which relates the center of light of the illuminated hemisphere to the 
center of mass. This factor is dependent on $Q$ and shown in Horne \& Schneider
(1989) for the optically thin and in Schwope et al.~(1993) for the 
optically thick models. The differences between both modes are negligible as
far as this factor is concerned.
The computation of the curves (2) in the $(Q,i)-$plane makes use further 
of the 
Roche geometry (approximation of the spherical equivalent Roche-radius
as a function of $Q$ given by Eggleton (1983)) and Kepler's third law. 
We have assumed that the secondary is a main sequence star and
can be described by a ZAMS mass-radius relation for late-type stars. 
With  $M_2 = 0.146$\,M$_\odot$, the 
empirical relation by Caillault \& Patterson (1990) predicts the least
compact secondary for \hu, the other extreme with $M_2 = 0.206$\,M$_\odot$
is due to the relation by VandenBerg et al.~(1983) (solid and dashed lines in 
Fig.~\ref{qi_rel}). 
Relation (3) finally is the result
of a fit-per-eye to the location and curvature of the horizontal 
stream in the tomogram which relies on the
assumption that the center of light along the ridge of \he2 \ 
marks the single-particle trajectory. This results in the smallest estimates 
of $Q$ with $Q \simeq 2.5$.

It is clear from Fig.~\ref{qi_rel} (as well as from several previous 
Figures) that no unique solution exists presently in the framework of the
models presented here. 
Relations (1) and (2) suggest high values of the mass ratio and the  
inclination with $Q \simeq 4.5 - 5.5, i > 85\degr$ for the Helium lines 
and $Q = 5 - 5.5, i > 87\degr$ for the 
Balmer lines (the latter are not shown in Fig.~\ref{qi_rel}).
The curvature of the stream of \he2, \h1\ and H$\beta$ on the other 
hand suggests, that the mass ratio might be as low as $Q = 2.5$ and that 
the inclination correspondingly might be as low as $i \simeq 80\degr$.
Both estimates of the mass ratio seem to exclude each other presently and
it appears likely that neither the NEL (in combination with $K_2$-correction
and ZAMS-approximation) nor the HVC (fitted by a single-particle trajectory)
can be used straightly to give an accurate mass estimate. 

The disagreement between both approaches becomes smaller if one assumes that
NEL emission is more concentrated towards
the center of mass of the secondary than we have assumed. An absorbing cloud 
of matter in the $L_1$-region which would suppress emission from the `nose'
of the secondary could e.g.~provide the necessary bias of the effective radial
velocity amplitude towards higher values (and could also lower the width of
the line profiles, Fig.~\ref{fwhm_nel}). 
The expected re-emission might be responsible for the longish 
extension of the bright spot near the $L_1$-point in the tomogram of H$\beta$.
That such bias exists appears likely from the observed higher NEL-velocity
of the Hydrogen Balmer lines with respect to the Helium lines. The `nose'
is more effectively shielded in the more optically thick Hydrogen lines.

The best way to solve the conflict and to calibrate the
M/R-relation of the secondary star would be 
to measure the radial velocity of the secondary star independently using
e.g.~photospheric absorption lines. This will help to sort out which 
feature (NEL, HVC) of which line may serve as {\it the} estimator of the
mass ratio.
It requires \hu\ to be observed
in a low state of accretion. Such a measurement is highly demanded in order to 
calibrate either of the methods used here for mass determination.
Another hint to the likely mass of the stars and the mass ratio could be 
derived from ingress/egress length into/from eclipse of the white dwarf
itself, rather than ingress/egress of the accretion spot on it.

Our model with $Q = 2.5$ implies that the secondary star 
in \hu\ must be as massive as 0.35\,M$_\odot$. It would be much 
more compact than a ZAMS-star and in consequence much more 
luminous. This would imply that the distance estimates given by 
Glenn et al.~(1994) and Schwope et al.~(1993) are lower limits only.

\medskip
We have presented high-resolution spectral observations of the eclipsing 
AM Herculis star \hu\ obtained when the system was in a high state of
accretion. We have identified three emission line components in our 
trailed spectra of \he2, \h1, and the H-Balmer lines. These could be
uniquely identified as originating on the illuminated hemisphere of the 
secondary star (the narrow emission line NEL), the stream in the orbital
plane (the high-velocity component HVC) following a ballistic trajectory,
and the stream coupled onto field lines out of the orbital plane 
(the broad-base component BBC). We have uniquely identified
the horizontal stream in this AM Her-binary by means of Doppler tomography
and could trace it from the $L_1$-point down to the stagnation region. 
On the assumption that the 
NEL of \he2 is formed on the surface of the Roche-lobe of the companion 
star we have determined its mass. This estimate is in disagreement with an 
estimate derived from the tomogram. We propose that Doppler tomography 
of the ballistic stream can be used as a new tool for the mass determination 
of accreting binaries provided we can calibrate this method using an
independent mass determination.

\acknowledgements
We thank M.~Giampapa for helpful comments concerning chromospheric emission 
from late-type stars and H.~Ritter for useful hints concerning their
mass-radius relation. We thank an anonymous referee for helpful comments.
This work was supported in part by the Deutsche Forschungsgemeinschaft
under grant Schw 536/4-1 and the BMFT under grant 50 OR 9403 5.

\end{document}